\documentstyle[graphicx]{mn}
\title{A {\tt SAURON} study of M32: measuring the intrinsic flattening and the central
black hole mass}
\author[E. K. Verolme et al.]
	{E. K. Verolme,$^1$\thanks{E-mail: verolme@strw.leidenuniv.nl}
	M. Cappellari,$^1$\thanks{European Space Agency external fellow}
	Y. Copin,$^2$	
	R.P. van der Marel,$^3$
	R. Bacon,$^4$
\newauthor M. Bureau,$^{1,5}$\thanks{Hubble fellow}
	R.L. Davies,$^6$
	B.M. Miller,$^7$
	P.T. de Zeeuw$^1$\\
$^1$Sterrewacht Leiden, Postbus 9513, 2300 RA Leiden, The Netherlands \\
$^2$Institut de Physique Nucl\'eaire de Lyon,  69622 Villeurbanne, France\\
$^3$Space Telescope Science Institute, 3700 San Martin Drive, Baltimore, MD
    21218, USA\\
$^4$CRAL-Observatoire, 9 Avenue Charles-Andr\'e, 69230 Saint-Genis-Laval,
    France\\
$^5$Columbia Astrophysics Laboratory, 550 West 120th Street, 1027 Pupin Hall,
    Mail Code 5247, New York 10027, NY, USA\\
$^6$Physics Department, University of Durham, South Road, Durham DH1 3LE, 
    United Kingdom\\
$^7$Gemini Observatory, Casilla 603, La Serena, Chile\\
}
\date{Accepted 0000 0000. Received 0000 0000; in original form 0000 0000}
\pubyear{0000}
\begin{document}
\maketitle

\begin{abstract} 
We present dynamical models of the nearby compact elliptical galaxy
M32, using high quality kinematical measurements, obtained with the
integral-field spectrograph {\tt SAURON} mounted on the William
Herschel Telescope on La Palma. We also include {\tt STIS} data obtained 
by Joseph et al. We find a best-fit black hole mass of $M_\bullet = 
(2.5 \pm 0.5) \times 10^6 M_\odot$ and a stellar $I$-band mass-to-light 
ratio of $(1.85 \pm 0.15)\,M_\odot/L_\odot$. For the first time, we are 
also able to constrain the inclination along which M32 is observed 
to $70^\circ \pm 5^\circ$. Assuming that M32 is indeed axisymmetric,
the averaged observed flattening of 0.73 then corresponds to an intrinsic 
flattening of $0.68 \pm 0.03$. 

These tight constraints are mainly caused by the use of integral-field data. 
We show this quantitatively by comparing with models that are constrained by 
multiple slits only. We show the phase-space distribution and intrinsic 
velocity structure of the best-fit model and investigate the effect of 
regularisation on the orbit distribution. 
\end{abstract}


\section{Introduction}

M32 is a high-surface brightness, compact E3 companion of the
Andromeda galaxy.  Ground-based kinematic measurements of this galaxy
(Tonry 1984, 1987) already showed a steep gradient in the central
velocity profile and a central dispersion peak, suggesting the presence of a
central compact object, presumably a supermassive black hole. Since
then, (spectroscopic) observations with increasing spatial resolution, both
ground-based (Dressler \& Richstone 1988; van der Marel et al. 1994a; Bender,
Kormendy \& Dehnen 1996) and with HST (Lauer et al. 1992; 
van der Marel, de Zeeuw \& Rix 1997; Joseph et al.\ 2001), strengthened the 
case for this black hole.

Simultaneously, dynamical models of ever improving quality were
built to match these data sets (Richstone, Bower \& Dressler
1990; van der Marel et al. 1994b; Qian et al.\ 1995; Dehnen 1995). For most
galaxies, the value of the best-fit black hole mass that is found
depends on the assumptions that are made about the distribution 
function (DF) of the galaxy. Two-integral models, which assume that 
the DF depends only on the two classical integrals of motion 
(the energy $E$ and the vertical component of the angular momentum 
$L_z$), generally tend to overpredict the central black hole mass, 
since they cannot provide sufficient radial motion (Magorrian et 
al.\ 1998; Gebhardt et al.\ 2000; Bower et al. 2001). 

The current state-of-the art models allow for the maximal degree of 
anisotropy by assuming that the distribution function depends on 
three integrals of motion (van der Marel et al.\ 1998, hereafter vdM98; 
Cretton et al.\ 1999, hereafter C99; Gebhardt et al. 2000, 2001). 
The internal kinematical structure 
of the best-fitting three-integral model of M32 closely resembles that of 
an $f(E,L_z)$ model. This explains why the central black hole mass that was 
found in M32 by the early models does not differ much from the current value 
of $(3.4 \pm 0.7)\times 10^6 M_\odot$ (vdM98).
 
This means that the mass of the central black hole and intrinsic
kinematical structure of M32 are constrained rather well. The main
uncertainty that remains is the inclination along which M32 is
observed, or, equivalently, the intrinsic flattening: edge-on models
to a data-set consisting of {\tt FOS} kinematics and ground-based
observations along four position angles fit equally well as models 
that are projected over $55^\circ$ (vdM98). When we combine this with the 
observed flattening of M32 (which is almost constant and equal to 
$0.73$ inside the central ten arcseconds, vdM98), we see that the intrinsic 
flattening of M32 is not very tightly constrained and can have any 
value between $0.55$ and $0.73$.

Recent developments in instrument design offer a way out: integral-field
spectrographs capture the full two-dimensional behaviour of objects
and are therefore expected to better constrain parameters such as the 
inclination. This tighter constraint on the intrinsic 
parameters can be explained from the behaviour of two-integral models, 
even though these generally provide less accurate fits to observational data. 
The part of a two-integral distribution function that is even in the 
velocities is only determined once the meridional plane density distribution 
is known (Qian et al.\ 1995). The full intrinsic velocity map is needed to 
constrain the odd part of the DF, which means we need to measure the 
line-of-sight velocity $v_{\rm los}$. Whereas these two lowest order 
velocity moments are sufficient to determine 
the {\it stellar} DF of a galaxy, the dark matter distribution is only 
constrained once the velocity anisotropy is known (Dejonghe 1987; 
Gerhard 1991, 1993). This quantity can be 
measured through the higher order velocity moments. It is therefore necessary 
to determine the full two-dimensional kinematic behaviour of a galaxy. 
Capturing this information with only a few slit positions, as is generally 
attempted, is sometimes possible, but not always, which implies that the 
distribution function remains unconstrained. 

These effects are even more important for {\it three}-integral
distribution functions, which, in most cases, provide better fits to the 
observations and have more freedom. Therefore, when galaxy models are 
constrained by kinematic observations along a modest number of slits, the 
intrinsic structure may remain largely unconstrained. In this paper, we 
revisit M32 using two-dimensional kinematical maps obtained with the panoramic 
integral-field spectrograph {\tt SAURON} (Bacon et al.\ 2001, hereafter Paper 
I; de Zeeuw et al.\ 2002, hereafter Paper II) and high-resolution spectra 
obtained with {\tt STIS} on board HST (Joseph et al.\ 2001). We show that the 
use of integral-field data places very tight constraints on the central black 
hole mass and mass-to-light ratio of M32, but also, for the first time, on the 
inclination, or equivalently, the intrinsic flattening. 

The paper is organized as follows: in Section \ref{section2}, we give a brief
summary of the data that are used to constrain the dynamical models,
which are described in Section \ref{section3}. The results are presented in
Section \ref{section4} and we show in Section \ref{section5} that the use of
integral-field data is crucial for most of the results that we
obtained. The properties of the best-fitting model are described in
Section \ref{section6}, and we conclude with a discussion in Section
\ref{section7}.

Throughout the paper we assume a distance of $D=0.7$ Mpc 
(Welch et al.\ 1986). 
This choice does not influence our conclusions, but sets the scale of the models
in physical units. Specifically, lengths and masses scale as $D$, while
mass-to-light ratios scale as $D^{-1}$.

\section{Observations}
\label{section2}

We use two fully complementary data sets obtained with current
state-of-the-art instruments: the integral-field spectrograph {\tt
SAURON} mounted on WHT, and {\tt STIS} on board HST. Integral-field 
spectrographs 
provide the full kinematic picture in one consistent data-set, while 
the high-resolution instrument {\tt STIS} is ideal to probe the central 
kinematics of nearby objects. With this combined data-set, we expect 
to be able to determine the intrinsic properties of M32 with great accuracy.

\subsection{{\tt SAURON} observations}

{\tt SAURON} is mounted on the WHT on La Palma and is specifically 
designed to measure the kinematics and line-strengths of nearby 
ellipticals, lenticulars and spiral bulges. Details on the instrument 
and the data reduction process can be found in Paper I. The instrument 
has two observing modes: in the low-resolution mode, the field of view 
is $30 \times 41$ arcsec at a pixel size of $0\farcs94$, while the 
high-resolution mode, used to zoom in on objects in good seeing 
conditions, has a pixel size of $0\farcs28$, corresponding to a 
field-of-view of $9 \times 11$ arcsec. 

\begin{figure*}
\includegraphics[width=17.cm,trim=1.cm 0cm 0cm 0cm]{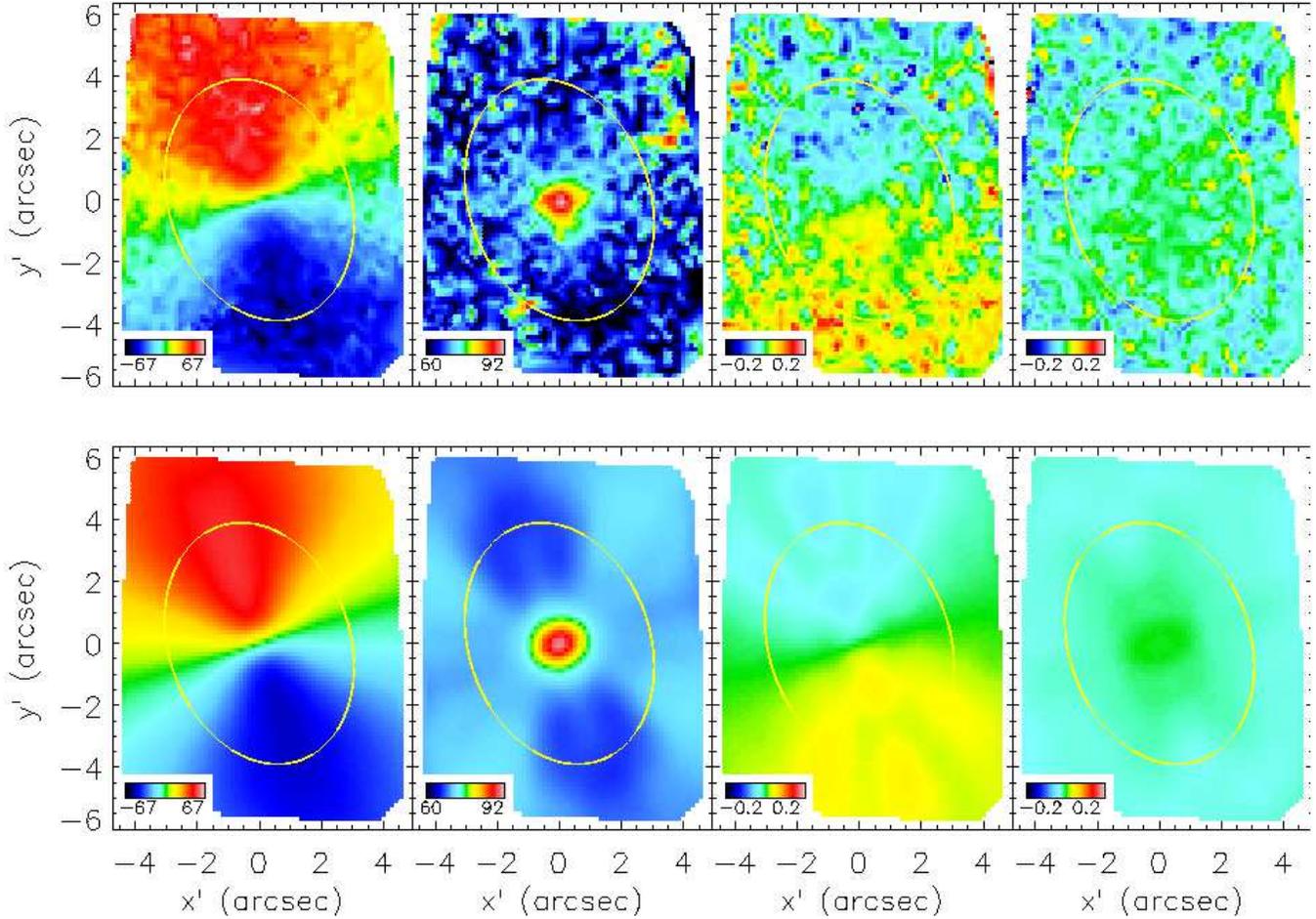}
\caption{\label{figure1} 
        {\bf Top}. 
	The kinematic maps of M32 (from left to right: the mean
	velocity, velocity dispersion and Gauss-Hermite moments
	$h_3$ and $h_4$), as observed with the integral-field 
	spectrograph {\tt SAURON}. The field of view is $9 \times 11$ arcsec. 
	{\bf Bottom}. 
	The best fit kinematic maps, obtained by adding the weighted 
	contributions of the best fitting set of orbits in the 
	{\tt SAURON} apertures.}
\end{figure*}

The {\tt SAURON} observations were made on October 15, 1999 (Paper
II). Relatively good seeing (FWHM$\approx 0\farcs95$) allowed us to
obtain the measurements in the high-resolution mode. The kinematical 
maps of M32, derived from a single 2700s exposure, are shown in the 
four upper panels of Fig.~\ref{figure1}. The velocity field reaches a 
peak value of $\approx$ 65 km\,s$^{-1}$; the misalignment between the 
rotation axis and the projected minor axis is less than two degrees, 
demonstrating that the measurements are consistent with axisymmetry. 
The dispersion of M32 drops well below the instrumental resolution 
($\sigma\approx$ 90 km\,s$^{-1}$) outside the inner few arcseconds. This 
results in a fourth-order Gauss-Hermite moment $h_4$ (measuring the symmetric 
deviations from a Gaussian, van der Marel \& Franx 1993) that is difficult to
measure. As is apparent from the fourth panel of Fig.~\ref{figure1}, it
is almost constant over the field.

A comparison of the {\tt SAURON} measurements with ground-based
long-slit data, obtained with the {\tt ISIS} spectrograph on WHT,
shows that the velocities differ only by $2.3 \pm 2.2$ km\,s$^{-1}$, 
the dispersions by $6.0 \pm 2.6$ km\,s$^{-1}$ and $h_3$ by 
$-0.009 \pm 0.016$ (Paper II; the accuracy of $h_4$ is difficult to 
determine). This indicates that there is no systematic offset in $V$
and $h_3$, while the disagreement between the dispersion measurements
is only significant at the 2$\sigma$-level. We do not correct for this 
and use all {\tt SAURON} measurements to constrain the models. 

\subsection{{\tt STIS} observations}

\begin{figure}
\includegraphics[width=8.cm,trim=0.cm 0cm 0cm 0cm]{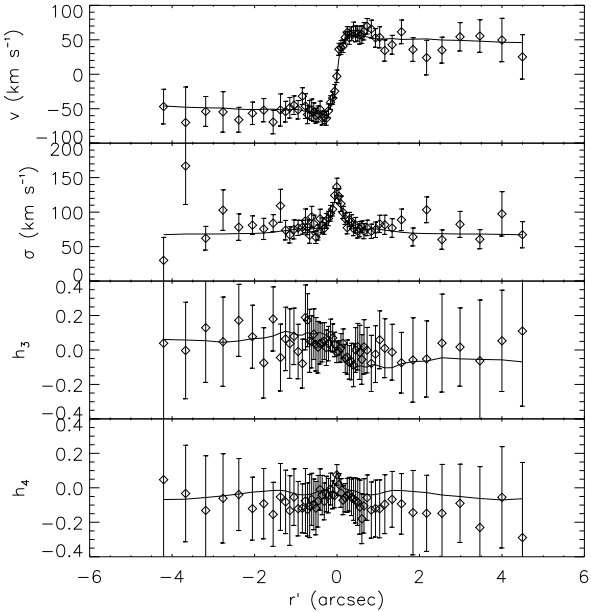}
\caption{\label{figure2} 
        The panels show the major axis velocity, dispersion,
        third and fourth order Gauss-Hermite moments of M32 
	obtained with the high-resolution spectrograph {\tt STIS}
	on board HST (diamonds, Joseph et al. 2001), together with 
	the predictions at the {\tt STIS} apertures of our best-fit 
	model (solid line).}
\end{figure}

The {\tt STIS} measurements (Fig.~\ref{figure2}) were obtained 
by Joseph et al.\ (2001), using the FCQ-method (Bender 1990). They
clearly show the main signatures of a central black hole: a steep velocity 
gradient and a peak in the dispersion profile near the center. 
By assuming that M32 has a
two-integral distribution function and by solving the Jeans equations,
Joseph et al.\ estimate that a central black hole with a mass of
$(2-4)\times 10^6 M_\odot$ best reproduces these signatures, 
in full harmony with the earlier results by vdM98. {\tt STIS} provides
a continuous spatial sampling at $\approx 0\farcs051$ pixels, avoiding
positioning problems that may occur when using aperture spectrographs
such as {\tt FOS} (van der Marel, de Zeeuw \& Rix 1997). Comparisons 
between these two
types of data are therefore only possible if such uncertainties are
taken into account (Joseph et al.\ 2001). The {\tt STIS} measurements
are superior to the {\tt FOS} data, since the spectral resolution and 
signal-to-noise ratio are much higher, an advantage when measuring 
the relatively low velocity and dispersion of M32. Because of this, we can
use up to the fourth order Gauss-Hermite moment, while {\tt FOS} was
only able to measure the first two.

\section{Method}
\label{section3}
\subsection{Mass models}

We construct dynamical models, based on Schwarzschild's
orbit superposition method (Schwarzschild 1979; 1982). We assume that 
the galaxy is axisymmetric and allow for the full degree of anisotropy 
by assuming that the distribution function depends on three integrals 
of motion. The implementation that is used here was originally developed 
by Rix et al.~(1997), vdM98 and C99, and was designed for use on
one-dimensional surface-brightness profiles (but see Cretton \& van
den Bosch 1999).  However, for M32, as for most nearby ellipticals,
high-quality two-dimensional imaging is available, so that more
sophisticated density parametrisations are possible. An example of
such a formalism is Multi-Gaussian Expansion (MGE, Monnet, Bacon \& Emsellem
1992; Emsellem, Monnet \& Bacon 1994), in which the set of Gaussians that 
best fits an image of the galaxy is found. The advantages of this formalism are 
that both axisymmetric and triaxial galaxies can be 
reproduced in a straightforward manner, the deprojection is 
simple and can be done analytically (under certain assumptions) once the
viewing angles are known, and convolution with (MGE) point spread functions 
is fast and simple. We have therefore adapted the axisymmetric Schwarzschild 
software to use MGE; the new algorithm was tested while modeling the 
dynamics of the counter-rotating core galaxy IC1459 (Cappellari et al.~2002). 
These tests have shown that the modeling results do not depend on the choice
of mass model: the best-fitting parameters and intrinsic (phase space)
structure of models that start from an MGE parameterisation with constant
axial ratio are identical to those of models that use a power-law density 
profile. 

An MGE parameterisation of M32 was obtained using the software of 
Cappellari (2002), by fitting to a combination of a high-resolution 
HST/WFPC2/F814W image and a ground-based $I$-band image obtained by 
Peletier (1993; $0\farcs549$ pixels$^{-1}$ with a seeing of 
$1\farcs27$) with the Isaac Newton Telescope on La Palma. Since 
we are constructing axisymmetric models, we do not allow for shifts in the 
position angles of the Gaussians and fix the centers 
of all individual Gaussians to the center of the galaxy. 
Each Gaussian therefore has three free parameters: the 
amplitude $P'$ (in units of $L_\odot\,{\rm pc}^{-2}$), the width $\sigma'$ in 
pixels (or arcseconds) and the projected flattening $q'$.
Eleven Gaussians
were used in the fit, since, upon adding more Gaussians, 
the RMS error of the fit does not change by more than 1\%. 
The parameters of these eleven Gaussians are listed in Table~\ref{table1}; 
this superposition reproduces the M32 surface brightness with an RMS 
error of $\approx$2.2 per cent (Cappellari 2002, Figure 6). This
is within the measurement errors. 
\looseness=-2

\begin{center}
\begin{table}
\caption{\label{table1} 
        The parameters of the eleven Gaussians in the MGE-fit to the 
	combined ground-based and HST/WFPC2/F814W image of M32. 
	The second column shows the amplitude $P'$, the third column
	shows the width $\sigma$ and the fourth the projected flattening 
	of each Gaussian.}
\begin{tabular}{c c c c}
\hline
number & $P' (L_{\odot}\,{\rm pc}^{-2})$ &$\sigma'$ (arcsec) & $q'$\\
		1 & 615412. &  0.0364   &  0.740\\
		2 & 517371. &  0.131 &  0.809\\
		3 & 324569. &  0.348  & 0.775\\
		4 & 175920. &  0.753  & 0.732\\
		5 & 54346  &  1.51   & 0.742\\
		6 & 22279 &  3.18   & 0.708\\
		7 & 11428 &  6.05   & 0.746\\
		8 & 4037.5 &  11.7    & 0.683\\
		9 & 2506.2 &  19.2    & 0.819\\
		10 & 1142.4 &  33.5   & 0.820\\
		11 & 226.4 &  95.8   & 0.827\\
\hline
\end{tabular}
\end{table}
\end{center}

\subsection{Axisymmetric three-integral Schwarzschild models}
\label{models}

We use the MGE-parameterisation to calculate axisymmetric
three-integral models with the modified Schwarzschild method.  Our
models are different from those of previous authors: we use 
independent kinematic data sets and we parameterise the mass density 
differently. These differences provide a useful test: if we
find best-fitting parameters that match the ones found by
e.g. vdM98, it means that both approaches are reliable and the
results are robust.

The set of best-fitting Gaussians is deprojected by assuming that the
galaxy is axisymmetric and choosing a value of the inclination\footnote{
The flattest Gaussian in the MGE superposition, in our case the eighth 
Gaussian, limits the range of possible inclinations to $i \geq 47^\circ$. 
Since this range includes values of $i$ that can be ruled out for other
physical reasons (see below), this mathematical limitation 
is not a problem.}. The resulting intrinsic luminosity density is 
multiplied by a constant mass-to-light ratio $M/L$ to obtain the 
intrinsic mass density. The stellar potential can then be calculated by 
applying Poisson's equation. A central supermassive black hole is mimicked 
by adding the potential of a central point mass $M_\bullet$. In the 
combined stellar and black hole potential, a representative orbit 
library is found by sampling the two classical integrals of motion, 
the energy $E$ and the vertical component of the angular momentum $L_z$,
together with the third integral $I_3$. 

Here, we only give a brief outline of the scheme that is used to sample 
these integrals; a more thorough description can be found in
C99. The orbital energy is sampled through a logarithmic grid in the 
radius $R_c$ of the circular orbit at energy $E$. This grid is chosen to
include more than $99.99$ per cent of the mass. While some of the orbits that
make up the missing mass might contribute to the density {\it in} our radial 
range, this contribution will be smaller than $0.01$ per cent and can 
therefore be safely ignored. 

At each energy, $L_z$ is sampled by a linear grid in the fractional 
angular momentum $L_z/L_{\rm max} \in \langle-1,1\rangle$, 
with $L_{\rm max}$ the maximum orbital angular momentum at energy $E$.
This grid does not include $L_z/L_{\rm max}=\{0,1\}$ to avoid problems with 
the numerical integration. This is not a severe limitation, as the 
orbits that are missed in this manner carry only a small fraction
of the mass in an axisymmetric model. The third integral 
is parameterised by the orbital starting point on the zero velocity curve. 

Every combination of the integrals $(E,L_z,I_3)$ defines a
separate stellar orbit. Since the third integral is generally not
known analytically, the orbital properties can only be obtained by
solving the equations of motion numerically. At every time step of
this orbit integration, which is carried out for a fixed number of
characteristic orbital periods (usually chosen to be the period of the
circular orbit at the given energy), the orbital energy is randomized
by drawing a number of arbitrary values from the energy bin around 
$E$, which are then assigned to the orbit. This 'dithering' of 
integral space results in
smoother orbital superpositions, which generally provide better fits
to the data (Zhao 1996; C99). 

The orbital superposition can be made even smoother by including the
so-called two-integral components (TICs, C99; Verolme \& de Zeeuw
2002) in the orbit library. These TICs are building blocks that obey
only the two classical integrals of motion $E$ and $L_z$ and
implicitly include stochastic orbits at the given values of energy
and angular momentum. Since they are smoother than most orbits, the
orbital superposition also suffers less from discretisation 
when they are included in the library. Here, we calculated model fits to
different data-sets (both long-slit and integral-field data) with and
without TICs in the orbit library, and found that the fit improves
only marginally upon inclusion of the TICs (RMS error of the fit
changes by less than 0.1\%). On the other hand, calculating observables
for the TICs at every aperture can be very time-consuming when many
data points are included in the fit, as is the case when using
integral-field data. To save computation time, we therefore decided
not to include the TICs in the derivation of the best-fit parameters.

The observables of all orbits (and TICs) in the library are stored on
grids that are adapted to the resolution of the observations. Using
these, the orbital superposition that best fits the data is found by
applying the non-negative least-squares (NNLS) routine of Lawson \&
Hanson (1974).  The (unavoidable) discrepancy between model
predictions and data, caused by noise in the data, discretisation 
effects and wrong choices of the model parameters, is expressed in 
a value for the $\chi^2$, defined as
\begin{eqnarray}
\chi^{2}=\sum_{i=1}^{N_d}\left(\frac{D^*_i-D_i}{\Delta D_i}\right)^{2},
\end{eqnarray}
where $N_d$ is the number of constraint points, $D_i$ is the 
observational constraint at the $i$-th data
point, $D^*_i$ is the model prediction at that point and $\Delta D_i$
is the uncertainty that is associated with this value (usually the
observational error). By varying the inclination, central black hole
mass and stellar mass-to-light ratio, we investigate which combination
$(M_\bullet, M/L,i)$ results in the overall smallest $\chi^2$ (the
overall best fit to the data). The value of $\chi^2$ for a single
model is of limited value, since the true number of degrees of freedom
is generally not known, but the difference in $\chi^2$ between a model
and the overall minimum value, $\Delta \chi^2 = \chi^2 - \chi^2_{\rm
min}$, is statistically meaningful (see Press et al.\ 1992). By using
the central limit theorem, it can be shown that this interpretation is
even valid when the errors $\Delta D_i$ are not Gaussian distributed 
(van der Marel et al. 2000), which is important when applying the 
statistic to models of observational data. This means
that we can assign the usual confidence levels to the $\Delta \chi^2$
distribution and determine the probability that a given set of
model parameters will occur.

Introducing integral-field data in numerical models also means
a considerable increase in the number of measurements that has to be 
reproduced: even when only one {\tt SAURON} pointing is used, a fit
to both the intrinsic and projected mass density and to all four
Gauss-Hermite moments results in approximately 8000 constraints.
Standard mathematical practice in minimization problems is to 
use more free parameters than constraints, which means we  
would need $\sim 10^4$ orbits to fit a {\tt SAURON} field. 
On the other hand, the behaviour of models that are only constrained 
by observational data is very similar to that of models with 
additional regularisation constraints (R97; C99; vdM98; Fig.~\ref{figure5} 
of this paper), even though regularised models always have more 
constraints than orbits (\S\ref{sec42}).

We conclude from this that it is not necessary to use more orbits than
constraints to obtain accurate and unbiased fits to the data. 
In our case, we decided to use an orbit library with
$N_o=20 \times 7 \times 7$ orbits. Since the orbital $L_z$ can be both positive 
or negative, depending on the chosen direction of rotation around the 
symmetry axis, this results in a library of 1960 orbits, implying that 
we have almost four constraints per orbit. 

When using integral-field data, the integration time of 
individual orbits may become extremely lengthy, as their properties have 
to be stored in many apertures. This is another reason not to use very large
orbit libraries. Integrating all orbits in our library requires about 
eight hours of computation time on a 1GHz machine with 
512Mb memory, while one NNLS fit takes approximately one hour on the 
same machine.   

\section{Results}
\label{section4}
\subsection{Best-fit intrinsic parameters}
 
We sample the model parameters $(M_\bullet, M/L, i)$ on a grid of 990 models, 
defined by fifteen values of the central black hole mass, spaced
linearly between $M_\bullet = 0$ and $M_\bullet = 4.5 \times 10^6 M_\odot$, 
eleven values of $M/L$ between 1.5 and 2.5, and inclinations $i=90^{\circ},
75^{\circ}, 70^{\circ}, 65^{\circ}, 60^{\circ}$ and $50^{\circ}$.
We do not consider smaller values of the inclination, as this would make
M32 intrinsically flatter than E7, i.e., flatter than any observed
elliptical galaxy. Models with equal ratios between the mass-to-light ratio 
and the central black hole mass differ only by a scaling factor in the 
velocity. This means that, for every inclination, only fifteen orbit libraries 
have to be calculated, each with a different ratio $M_\bullet/(M/L)$. Any 
point along a line of constant $M_\bullet/(M/L)$ can then be obtained 
by scaling the velocity prior to the NNLS fit. As a result, 
one full $\chi^2(M_\bullet,M/L)$ grid (at fixed inclination) takes about 
twelve days of computation time.

\begin{figure}
\includegraphics[width=8.cm,trim=0.cm 0cm 0cm 0cm]{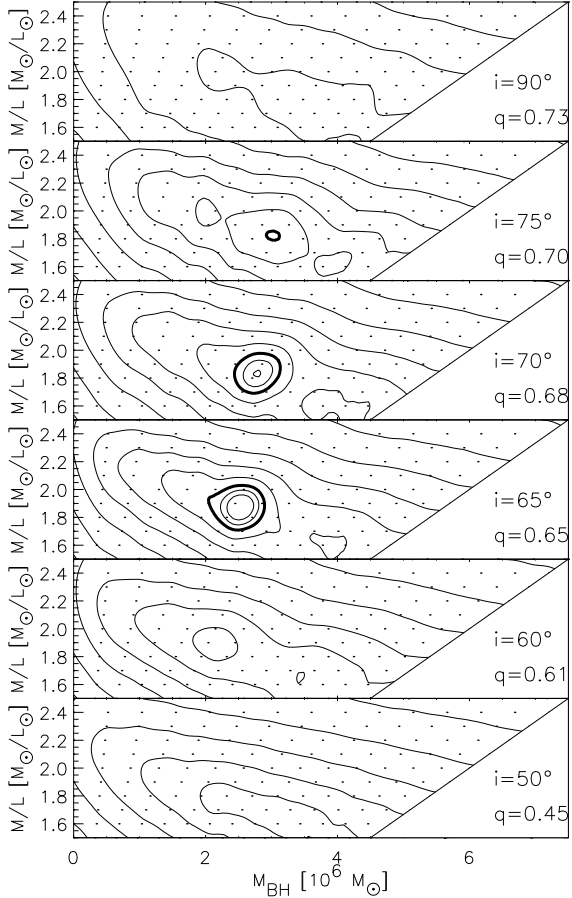}
\caption{\label{figure3} 
        $\Delta\chi^2$ as a function of central	black hole mass 
	$M_\bullet$, stellar mass-to-light ratio $M/L$
	and inclination $i$. The inclination varies from $90^{\circ}$
	(top panel) to $50^{\circ}$ (bottom panel), and 
	$\chi^2$ is offset such that the overall minimum is zero.  
	Contours are drawn at the formal confidence levels for a
	$\Delta\chi^2$-distribution with three degrees of freedom,
	with the inner three contours corresponding to the 1$\sigma$,
	2$\sigma$ and 3$\sigma$ confidence levels (the thick contour
	indicates the 3$\sigma$ level); subsequent contours correspond
	to a factor of two increase in $\Delta \chi^2$.}
\end{figure}

For each of the 990 models that were obtained in this manner, we calculated 
the fit to the combined {\tt SAURON} and {\tt STIS} data. Contour plots of 
the values of $\chi^2 (M_\bullet,M/L,i)$, offset such that the 
overall minimum value is zero, are shown in Fig.~\ref{figure3}. 
The contour levels 
correspond to the confidence levels of a $\Delta \chi^2$ distribution 
with {\it three} degrees of freedom (with the thick contour in the 
figure indicating the 3$\sigma$ confidence level), so that the volume in which 
the intrinsic parameters of M32 are most likely located can be readily 
assessed. The shape of the contours is very regular, even though no smoothing
has been applied.

The best-fitting {\it edge-on} model has a black hole mass of 
$(4.0 \pm 0.5) \times 10^6 M_\odot$ (3$\sigma$-error), which agrees
very well with the vdM98's value of $(3.4 \pm 0.7) \times 10^6 M_\odot$, even 
though the observational data sets are independent. However, we see from 
Fig.~\ref{figure3} that models with an inclination in the range $i=70^\circ \pm
5^\circ$ provide significantly better fits to the data than models with other
inclinations, which are ruled out at high confidence. This narrow 
constraint on the inclination corresponds to an intrinsic flattening of 
approximately $0.68 \pm 0.03$ (given the observed flattening of 0.73). 

The overall best-fit model, located at $i=70^\circ$, has a black hole mass of 
$(2.5 \pm 0.5) \times 10^6 M_\odot$ and an $I$-band mass-to-light ratio of 
$(1.85 \pm 0.15)\, M_\odot/L_{\odot}$. The fit of this model to the 
{\tt SAURON} data is shown in the bottom panels of Fig.~\ref{figure1}; 
the fit to the {\tt STIS} data is given in Fig.~\ref{figure2}.

\subsection{Applying regularisation}
\label{sec42}
The matrix problem that is solved by the NNLS routine is often
ill-conditioned, resulting in orbital weights $\gamma(E, L_z,I_3)$
that vary rapidly as a function of the integrals of motion. This can
be avoided by imposing regularisation constraints on the orbital
weights (see e.g. Press et al.\ 1992; Merritt 1993; Zhao 1996; 
Rix et al.\ 1997); another method, with similar effect, is to add
maximum entropy constraints (Richstone \& Tremaine 1988).
Applying regularisation forces the orbital weights towards a smooth function 
of $E, L_z$ and $I_3$ by minimizing the $n$-th order derivatives
$\partial^n\gamma(E,L_z,I_3)/\partial E^n,
\partial^n\gamma(E,L_z,I_3)/\partial L^n_z$ and
$\partial^n\gamma(E,L_z,I_3)/\partial I^n_3$. The degree of
smoothing is determined by the order $n$ and by the maximum
fractional error $\Delta$ that the derivatives are allowed to have.
For all regularised models that are described here, we have
minimized the second-order derivative ($n=2$).

\begin{figure}
\includegraphics[width=8.cm,trim=0.5cm 0cm 0.cm 0cm]{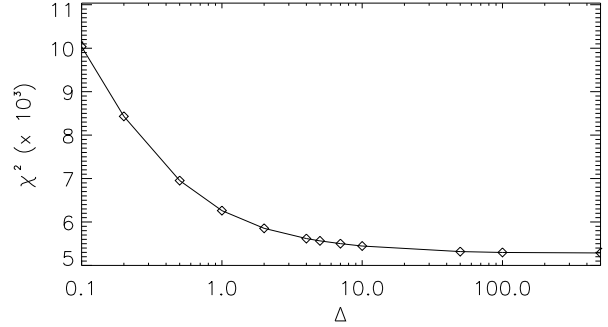}
\caption{\label{figure4} 
        $\chi^2$ as a function of the regularisation parameter, at 
	fixed $M_\bullet$, $M/L$ and inclination.  The model with no 
	regularisation is located at $\Delta=\infty$, but is placed at 
	$\Delta=500$ to include the value in the plot.}
\end{figure}

The optimal value of the regularisation error $\Delta$ was determined
by calculating model fits for a fixed combination of ($M_\bullet, M/L,
i$) and varying $\Delta$ between $0.1$ (high regularisation) and
$\infty$ (no regularisation). Fig.~\ref{figure4} shows that $\chi^2$
is large for small values of $\Delta$, but quickly decreases to an
almost constant level for $\Delta > 4$.
Since the ideal model both fits the constraints and has a smooth
distribution of orbital weights, we use $\Delta=4$ whenever
regularisation is needed. A similar degree of smoothing was used earlier 
in applications that use the same regularisation technique (vdM98; Cretton 
\& van den Bosch 1999) and in models that use the maximum entropy method 
(Gebhardt et al. 2000).

\begin{figure}
\includegraphics[width=8.cm,trim=0.5cm 0cm 0cm 0.cm]{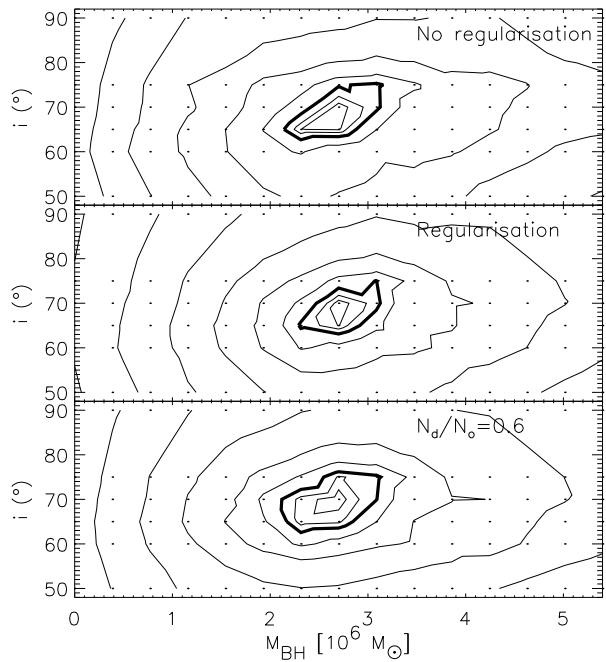}
\caption{\label{figure5} 
        $\Delta\chi^2$ as a function of central black
	hole mass and inclination, for a fixed $M/L$ of 1.8, for
	models without regularisation (top panel) and with
	regularisation (middle panel). The shape of the contours does
	not change upon adding regularisation. The bottom panel was calculated
	with $N_d/N_o=0.6$ and no regularisation, showing that our 
	results do not depend on the number of orbits in the fit.}
\end{figure}

Including smoothing constraints in the NNLS fit will, in
general, change the value of $\chi^2$ and therefore may affect the
shape of the contours in plots such as Fig.~\ref{figure3}. The top panel
of Fig.~\ref{figure5} shows a cross-section of Fig.~\ref{figure3} at
$M/L=1.8$; the $\chi^2$ of fits with the same parameters, but
including regularisation, is shown in the middle panel. Since the
shape of the contours does not change significantly, applying regularisation 
does not change the overall best-fit parameters, while it allows us to better 
study the phase-space structure and internal motions (see Fig.~\ref{figure7} 
below).

The two upper panels were calculated using four constraints per 
orbit ($N_d/N_o=4$, the value that is used throughout the paper). Although
we already argued in \S~\ref{models} that our results are not sensitive
to this parameter, we investigated this more quantitatively by
calculating (unregularised) models for which $N_d/N_o=0.6$, which means 
there are almost two orbits per constraint point. The result of this is 
shown in the bottom panel, which indeed illustrates that our results do not
depend on the number of orbits that is used. 

\section{The effect of integral-field data}
\label{section5}

At this point, although we have shown that the model parameters are
constrained much more tightly than in any previous model, it is not
entirely clear to what extent this is caused by the use of
integral-field data.  We also have to take into account that we have
used a very dense sampling of $M_\bullet$ and $M/L$, that we were able
to calculate models at more inclinations than vdM98, and that the {\tt
STIS} data is of higher quality than the {\tt FOS} observations. A
more quantitative estimate of the influence of the integral-field data
on the $\chi^2$ contours and on the best fitting parameters can be
obtained by repeating the procedure of Section \ref{section3} for a data set
that closely resembles the one used by vdM98. Since they fitted to a
combination of {\tt FOS} pointings with ground-based long-slit
observations along the major, minor, $45^\circ$ and $135^\circ$ axes 
(which agree with {\tt SAURON}), a similar data set can be
obtained by extracting virtual `slits' from the {\tt SAURON} field along the
same position angles and combining these with the {\tt STIS} data.

\begin{figure}
\includegraphics{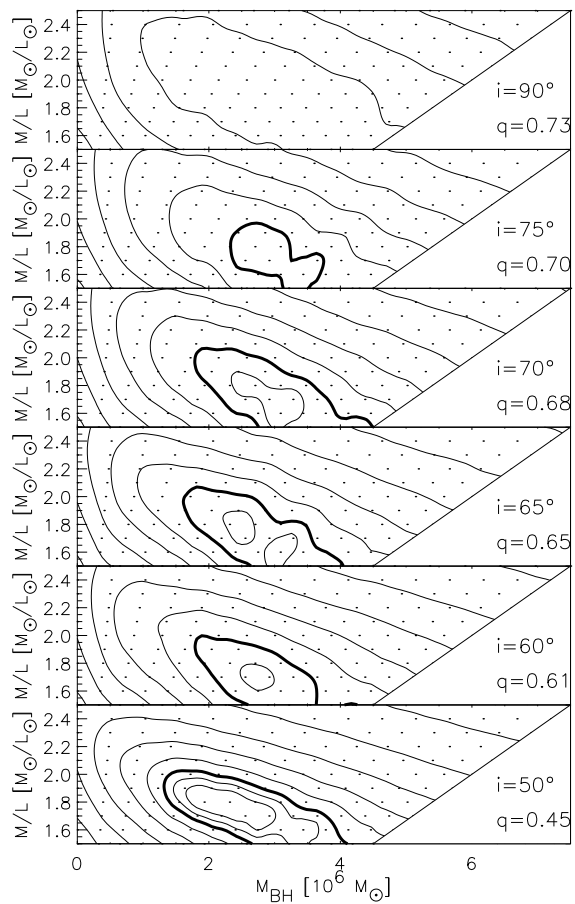}
\caption{\label{figure6} 
        Similar to Fig.~\ref{figure3}, for model fits
	to a data set consisting of the {\tt STIS} data and virtual
	`slits' simulated from the {\tt SAURON} data. The range of
	best-fitting values in all three model parameters is much
	larger than in Fig.~\ref{figure3}, which indicates that using
	integral-field data tightens the constraint on all model
	parameters.}
\end{figure}

This data set was used to constrain models with the same parameter ranges
as described in Section \ref{section3}. Fig.~\ref{figure6} shows the
$\chi^2$ of these fits.  The best-fitting black hole mass is
$M_\bullet = (2.8 \pm 1.2) \times 10^6 M_\odot$, the $I$-band mass-to-light 
ratio is $(1.75\pm 0.25)\, M_\odot/ L_{\odot}$ and the inclination can have any
value below $75^\circ$. A comparison with the best-fit parameters
of Section \ref{section4} indicates that the volume of intrinsic parameters 
providing a reasonable fit to the data is significantly larger for 
the slit data set than when the entire {\tt SAURON} field is used. We also 
notice that, although the use of integral-field data tightens the constraint
on {\it all} model parameters (with a fractional change in the allowed
range of $M_\bullet$ of 2.4 when changing from integral-field to slit
data, and 1.7 for $M/L$), the most dramatic effect occurs for the
inclination, which has an eight times larger allowed range when 
the galaxy is observed along a few slits only.

This can be explained from the behaviour of {\it two}-integral models, even 
though these generally provide less accurate fits to observational data.
As was mentioned in the Introduction, in order to constrain the even part of 
a two-integral distribution function, one needs to specify the meridional 
plane density distribution (Qian et al. 1995). The odd part is 
determined once the full intrinsic velocity field is known, which can only be
measured through the projected velocity field $v_{z'}(x',y')$ (in which $x'$
and $y'$ are the coordinates in the plane of the sky). Furthermore, even if the 
total {\it stellar} DF is determined accurately, the dark matter distribution 
is still unconstrained. This can be solved by measuring the higher order 
velocity moments, which are related to the anisotropy of the velocity 
distribution.

When the projected velocity moments are simple functions of $x'$ and $y'$, 
measurements at a small number of positions are sufficient to determine the 
behaviour at all intermediate position angles. Indeed, some very specific 
two-integral models, such as the power-law galaxies (Evans \& de Zeeuw 1994), 
have velocity fields that are quadratic in $x'$ and $y'$, implying that
observations along two perpendicular slits are sufficient to determine
their full two-dimensional kinematical behaviour. 

However, most galaxies require a distribution function that depends on 
{\it three} integrals of motion. The velocity fields of these three-integral
models (and, in fact, of most generally applicable two-integral
models) are usually much more complicated functions of the projected
coordinates (Dehnen \& Gerhard 1993; de Bruijne, van der Marel \& de Zeeuw 
1996). This means that these velocity
fields cannot be characterized by a (small) number of slit
orientations. It is therefore not surprising that realistic models
constrained by observations along only a few slit orientations
cannot provide much information about the intrinsic shape of the
galaxy. It also implies that integral-field data places constraints on
the intrinsic shapes of galaxies, as well as on their mass
distribution.

We see in both Fig.~\ref{figure3} and Fig.~\ref{figure6} that the
best-fitting black hole mass of each separate panel increases with
inclination. This behaviour can be understood at least partially from the 
following: the contribution of the intrinsic rotational velocity 
$v_\phi$ to the line-of-sight velocity decreases proportionally to $\sin i$. 
However, at smaller inclinations, the model must also be intrinsically flatter, 
which means $v_\phi$ is larger when lowering $i$. For a two-integral model, 
the increase in 
$v_\phi$ is dominant, so that there is more net velocity at smaller values 
of the inclination. This also implies that less mass is needed to obtain the 
same velocity field. Unfortunately, since there is a trade-off between the 
mass-to-light ratio and the central black hole mass and because we are dealing 
with {\it three}-integral models, it is not clear to what extent these 
considerations can explain the effect we observe. 

\section{Properties of the best-fit model}
\label{section6}

\begin{figure*}
\includegraphics[width=17.cm,height=10.cm,trim=1cm 0cm 0cm 0cm]{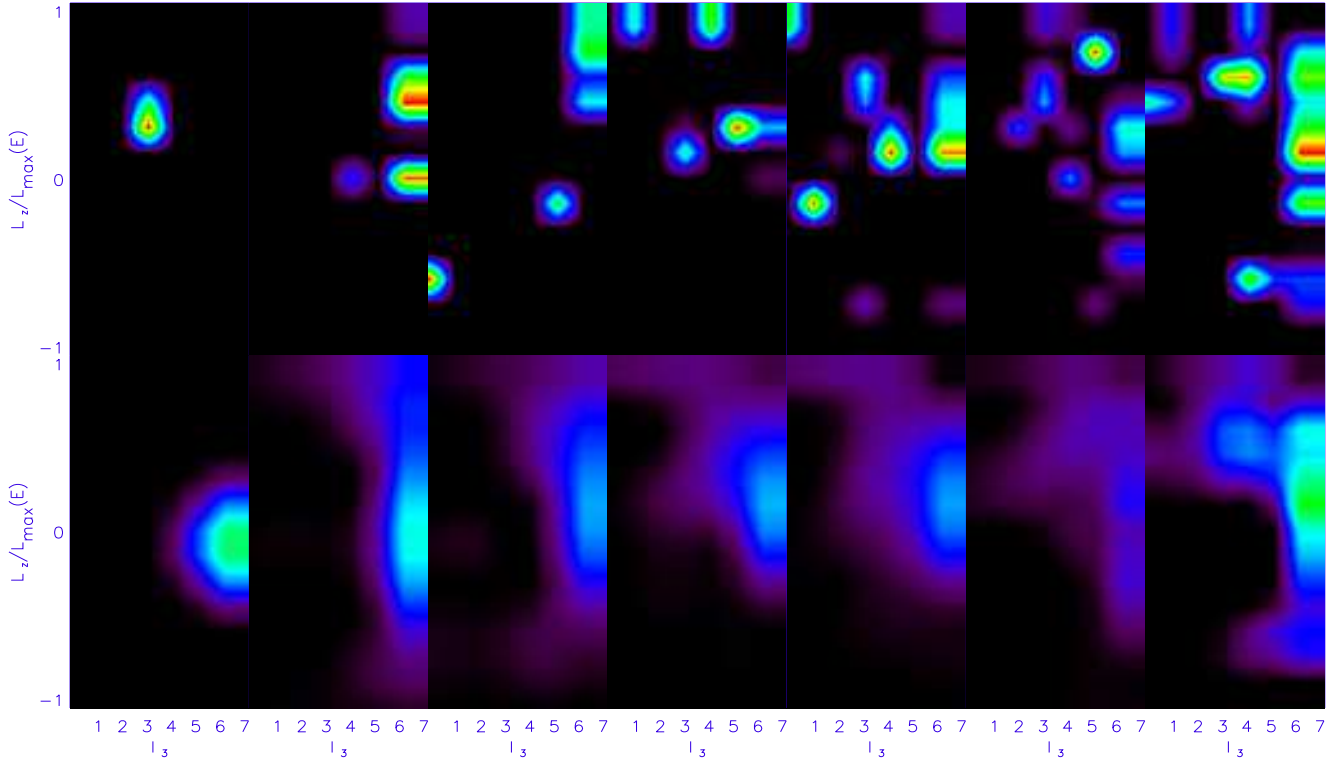}
\caption{\label{figure7} 
        Best-fit model orbital weights as a
	function of the three integrals of motion $E, L_z$ and
	$I_3$. Each panel shows the orbital weights as a function of
	$I_3$ and $L_z$ (in units of the angular momentum
	of the circular orbit; since orbits can move in both the
	prograde and retrograde direction, $L_z/L_{\rm max}$ can be
	either positive or negative) at	constant $E$. The energy 
	decreases from left to right,
	corresponding to an increasing mean orbital radius
	(from $0\farcs01$ in the left panel to $\approx 10$ arcsec in the
	right panel. Larger radii are not shown, since these are not
	meaningfully constrained by the data).  
	{\bf Top panels}: the rapidly varying orbital weights for 
	the unregularised model.
	{\bf Bottom panels}: including a modest degree of
	regularisation ($\Delta = 4$) smoothens the orbital weights,
	while still fitting the constraints. The color scheme is the same as
	in the upper panels.}
\end{figure*}
The upper panels of Fig.~\ref{figure7} show the phase-space distribution 
of the orbits of the non-regularised
best-fit model as a function of $I_3$ and $L_z$, for constant
$E$. The energy decreases from left to right, corresponding to an
increasing distance from the galaxy center. The bottom panels are similar, 
for the model with a regularisation error of $\Delta=4$.

\begin{figure}
\includegraphics[width=8.cm,trim=0.cm 0cm 0.cm 0.cm]{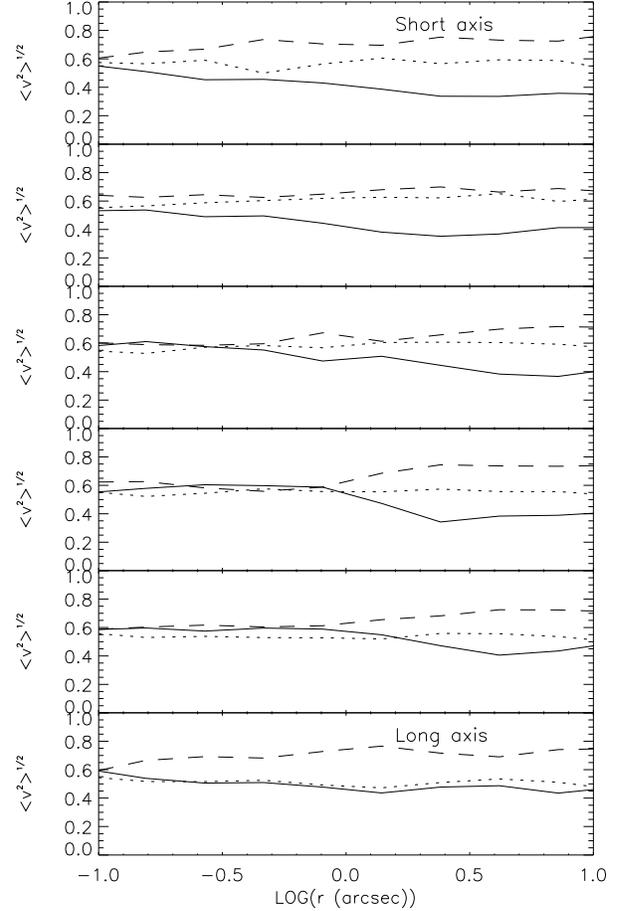}
\caption{\label{figure8} 
        Three intrinsic velocity moments $\langle
	v^2_r \rangle^{\frac{1}{2}}$ (solid line), $\langle v^2_\theta
	\rangle^{\frac{1}{2}}$ (dotted line) and $\langle v^2_\phi
	\rangle^{\frac{1}{2}}$ (dashed line) as a function of the
	meridional plane radius and the angle from the symmetry axis, 
	for the best-fitting model without regularisation. The velocity
	moments are normalized by the total RMS motion, $\langle
	v^2\rangle$. }
\end{figure}

We can study the intrinsic dynamical structure of the best-fitting
model ($M_\bullet=2.5 \times 10^6 M_\odot$, $M/L$=1.85 $M_\odot/L_{\odot,I}$,
$i=70^\circ$) by adding the appropriate moments of the orbits with
non-zero weight. This implies that we can study
the degree of anisotropy of the model, and from this determine to what
extent the galaxy resembles a two-integral model. Fig.~\ref{figure8}
shows the intrinsic velocity moments $\langle v^2_r
\rangle^{\frac{1}{2}}$ (solid line), $\langle v^2_\theta
\rangle^{\frac{1}{2}}$ (dotted line) and $\langle v^2_\phi
\rangle^{\frac{1}{2}}$ (dashed line) of the best-fitting model, as a
function of the meridional plane radius and the angle from the symmetry axis. 
The velocity moments are normalized by
the total RMS motion $\langle v^2 \rangle = \langle
v^2_r\rangle+\langle v^2_\theta\rangle+\langle v^2_\phi\rangle$.
We see that the model resembles an $f(E,L_z)$ model near the equatorial plane
($\langle v^2_r \rangle^{\frac{1}{2}}=\langle v^2_\theta
\rangle^ {\frac{1}{2}}$) and becomes increasingly more
anisotropic towards the symmetry axis. This indicates that a two-integral
model is not suitable to fit all constraints simultaneously, in
agreement with the findings of vdM98 (their Fig.~10). Furthermore, as 
in their models, azimuthal motion dominates.

For completeness, we have checked the internal kinematical structure
of the pseudo-slit model for the same parameters as were used for
Fig.~\ref{figure8}. Since this combination of intrinsic parameters is
inside the 3$\sigma$ contour of Fig.~\ref{figure6}, and therefore
also provides a good fit to the data, the results resemble those of
Fig.~\ref{figure8} closely. The main difference between using slits and 
integral-field data is that more combinations of intrinsic parameters 
can be ruled out by integral-field data. A point 
($M_\bullet, M/L, i$) that is not inside the 3$\sigma$ contour of 
Fig.~\ref{figure3}, but cannot be ruled out by
observations along a few slits, will therefore in general show a
different intrinsic kinematical structure.

\begin{figure}
\includegraphics[width=8.cm,trim=.5cm 0.cm 0.cm 0.cm]{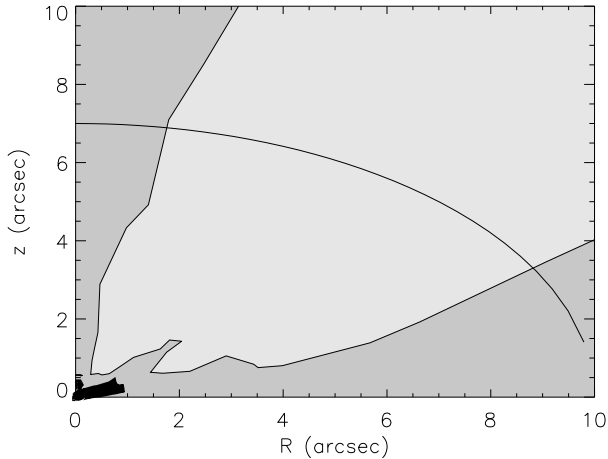}
\caption{\label{binney} 
        A greyscale plot of the anisotropy parameter $\beta_B$ as
	a function of the meridional plane coordinates $R$ and $z$,
	together with a typical isodensity contour (using
	$i=70^\circ$). The values were obtained by summing up the
	intrinsic moments of the set of orbits with non-zero weight in
	the best-fitting model. The darkest regions correspond to radial 
	anisotropy ($\beta_B>0$); the lightest region
        shows the points where the model is tangentially
	anisotropic ($\beta_B<-1$). In all other points, the 
	model is only mildly anisotropic ($-1 < \beta_B < 0$).}
\end{figure}

The anisotropy of a model can also be expressed in terms of an
anisotropy parameter. A possible definition of this parameter is 
\begin{eqnarray}
	\beta_B=1-\frac{\langle v^2_\theta \rangle^{\frac{1}{2}}+\langle 
                                             v^2_\phi
	\rangle^{\frac{1}{2}}}{2\,\langle v^2_r \rangle^{\frac{1}{2}}}.
\end{eqnarray}
When the model is fully isotropic, $\beta_B=0$, while $\beta_B=1$ for
a model that consists entirely of radial orbits, and $\beta_B=-\infty$
describes a model with only circular orbits. Fig.~\ref{binney} shows
this parameter as a function of the meridional plane coordinates $R$
and $z$, over the range that is constrained by the {\tt SAURON} data. A 
typical isodensity contour is overplotted (using $i=70^\circ$). We see
that $\beta_B$ is positive close to the galaxy center, so that the velocity
distribution is radially anisotropic there. In all other parts of the
meridional plane, the model is azimuthally anisotropic. This result does not
change upon adding regularisation.

Dynamical models of elliptical galaxies have revealed that a
few other objects are also tangentially anisotropic 
(NGC4697, Dejonghe et al. 1996; NGC1700, Statler et al. 1999), while
the majority seems to be radially anisotropic (M87, Merritt \& Oh 
1997; NGC2324, R97; NGC6703, Gerhard et al. 1998; M32, vdM98; 
NGC1600, Matthias \& 
Gerhard 1999; NGC3379, NGC3377, NGC4473, NGC5845, Gebhardt et al. 2000; 
NGC1399, Saglia et al. 2000). However, this radial anisotropy is usually rather small
and sometimes even accompanied by a transition to isotropy at small
radii. Furthermore, not all models included the full velocity profile 
in the fit, which implies that the mass-anisotropy degeneracy may not have 
been broken. Finally, our models are the first to use fully two-dimensional
kinematical observations, which makes comparison with earlier models even
harder. Therefore, although our results seem to contradict these
findings, we must also conclude that our knowledge of anisotropy is still limited. 

\vskip 1.0truecm

\section{Conclusions}
\label{section7}

We have presented dynamical models of the nearby compact elliptical
M32, using data from the integral-field spectrograph {\tt SAURON} and
from {\tt STIS} on board HST.  We have shown that our modeling
software is able to deal with two-dimensional kinematical information
and that the integral-field data tightens the constraint on all
intrinsic model parameters considerably.

The axisymmetric three-integral model that best fits the data has a
black hole mass of $(2.5 \pm 0.5) \times 10^6 M_\odot$ and a stellar
$I$-band mass-to-light ratio of $(1.85 \pm 0.15)\, M_\odot/L_{\odot}$.  
These values confirm the best-fit parameters that were obtained by previous
authors, although our modeling procedure is different: we use a fully
independent kinematical data-set and a different parameterisation 
for the mass density. Despite these differences in the methods, the 
best-fitting parameters that we find match the ones that were found by 
e.g. vdM98. This means that both approaches are reliable and that the
results are robust.

For the first time, we are able to determine the inclination along
which M32 is observed with great accuracy: the best-fit value is
$70^\circ \pm 5^\circ$.  If M32 is indeed axisymmetric,
the averaged observed flattening of 0.73 then corresponds to an 
intrinsic flattening of $0.68 \pm 0.03$. We have shown in Section 
\ref{section5} that 
this tight constraint is mainly caused by the use of integral-field data. This
implies that integral-field data will provide us with more insight
into the internal structure and kinematics of such objects. 

Although M32 is consistent with axisymmetry, it may be intrinsically triaxial,
but seen along one of its principal planes. Allowing for an intrinsically 
triaxial object also would enable us to study the effects of uncertainties 
in the deprojection on the best-fitting parameters more closely.
An extension of our method to triaxiality is in progress (Verolme et al.\ 
2002, in prep.), which will allow us to verify the assumptions 
made here. Furthermore, the tests that were carried out in this paper 
and the agreement with the results of other authors indicate that, 
given the assumptions, the results are robust.

We have obtained observations with {\tt SAURON} of a representative
sample of nearby ellipticals, lenticulars and spiral bulges, for most 
of which high-resolution {\tt STIS} data is, or will become, available. 
We are in the process of applying the axisymmetric version of 
Schwarzschild's method on {\tt SAURON} observations of a few of the 
sample galaxies that appear consistent with axisymmetry 
(e.g. NGC821, NGC3377, NGC2974).  The results of these dynamical models 
will provide us with the intrinsic parameters
of a considerable number of objects, giving us unique insight into the
formation and evolution of early-type galaxies.

\vspace{0.5cm}

\noindent{\bf Acknowledgements}\\ 
We thank Karl Gebhardt, Richard McDermid and Glenn van de Ven for
useful discussions, and Eric Emsellem, Harald Kuntschner and  
Reynier Peletier for a critical reading of the manuscript. YC 
acknowledges support through a European Community Marie Curie Fellowship.
MB acknowledges support from NASA through Hubble Fellowship grant
HST-HF-01136.01 awarded by the Space Telescope Science Institute, which is
operated by the Association of Universities for Research in Astronomy, Inc., 
for NASA, under contract NAS 5-26555.

\bsp
\label{lastpage}
\end{document}